\documentclass{aa}
\usepackage{graphicx}

\newcommand{\beq}{\begin{equation}}
\newcommand{\eeq}{\end{equation}}
\newcommand{\beqn}{\begin{eqnarray}}
\newcommand{\eeqn}{\end{eqnarray}}
\newcommand{\lppr}{\stackrel{<}{\scriptstyle \sim}}
\newcommand{\gppr}{\stackrel{>}{\scriptstyle \sim}}

\begin{document}
   \title{On the central black hole mass in Mkn~501}
  
   \author{F.M.~Rieger\inst{1,2} \and K.~Mannheim\inst{1}}
          
   \offprints{F.M.~Rieger, frieger@uni-sw.gwdg.de}

   \institute{Institut f\"ur Theoretische Physik und Astrophysik der
              Universit\"at W\"urzburg, Am Hubland, D-97074 W\"urzburg\\
               \email{frieger@uni-sw.gwdg.de; mannheim@astro.uni-wuerzburg.de}
              \and
              Universit\"ats-Sternwarte G\"ottingen, Geismarlandstr. 11,
              D-37083 G\"ottingen}

   \date{Received 17 June 2002; accepted 8 October 2002}

   \abstract{We analyse the apparent disagreement between the mass estimates 
   of the central black hole(s) in Mkn~501 based on (i) the observations of 
   the host galaxy, (ii) the high energy (HE) emission mechanism, and (iii) 
   the modulation of the beamed radiation by a black hole (BH) binary system.
   While method (i) seems to imply a central mass $\gppr 5\times 10^8\,
   M_{\odot}$, method (ii) suggests a BH mass less than $\simeq 
   6 \times 10^7\, M_{\odot}$. We critically discuss the estimates inferred 
   from (i) showing that current uncertainties may permit a central mass 
   as low as $\simeq (2-3)\times 10^8\,M_{\odot}$.
   We demonstrate that in this case the estimates (i) and (ii) might be 
   brought into agreement by assuming a binary BH system where the jet 
   dominating the HE emission originates from the less massive (secondary) 
   BH as suggested by method (iii). On the other hand, if Mkn~501 has in 
   fact a high central BH mass of order $10^9\, M_{\odot}$, a change of 
   fundamental assumptions seems to be required in the context of several 
   HE emission models. We show, that in this case a binary scenario following 
   (iii) may be still possible if the jet which dominates the emission
   emerges from the more massive (primary) BH and if the binary evolution 
   passes through phases of super-Eddington accretion and/or decreased 
   conversion efficiency.    
   }     
   \titlerunning{On the central black hole mass in Mkn~501}
   \maketitle


\section{Introduction}
    According to the commonly accepted paradigm, Active Galactic Nuclei 
    (AGN) are thought to harbour supermassive black holes (BH) 
    surrounded by geometrically thin accretion disks, the latter
    triggering the formation of relativistic jets. 
    Increasing evidence now indicates, however, that this picture 
    probably has to be expanded: 
    First, hierarchical galaxy evolution schemes suggest, that as 
    a result of mergers between galaxies, binary BH systems (BBHSs) should 
    be generally expected in the center of elliptical galaxies, and hence, 
    for example, in the typical hosts of BL Lac objects (cf. Begelman et 
    al.~1980; Kauffmann~1996; Richstone et al.~\cite{richstoneetal1998}; 
    Milosavljevic \& Merritt~\cite{milo01}; Yu~\cite{yu2002}).
    Secondly, from a phenomenological point of view, a multitude of 
    observational evidence has been plausibly related to the presence of 
    BBHSs in AGN, including the observed misalignment, precession and wiggling
    of extragalactic jets (e.g. Conway \& Wrobel~\cite{conwaywrobel1995}; 
    Kaastra \& Roos~\cite{kaastraroos1992}), periodic optical outburst events 
    as observed in the quasar OJ~287 (e.g. Sillanp\"a\"a et 
    al.~\cite{sillanetal1988}; Lehto \&
    Valtonen~\cite{letho1996}; Villata et al.~\cite{villataetal1998}) and the 
    helical motion of knots (e.g. Abraham \& Carrara~\cite{abrahamcarrara1998};
    Britzen et al.~\cite{britzenetal2001}).\\
        One expects that a binary framework for AGN may be particularly 
    relevant for our understanding of BL Lac objects such as Mkn~501 
    (e.g. De Paolis et al.~\cite{depaolisetal2002}).\\
        Mkn~501 ($z=0.034$) is one of at least four AGN which have been 
    reliably detected at TeV energies (see Catanese \& 
    Weekes~\cite{cataneseweekes99} for a review). From optical observations 
    its host is known to be the elliptical galaxy UGC 10599 (cf. Stickel et 
    al.~\cite{stickeletal93}). 
    As a BL Lac object, Mkn~501 belongs to the blazar class of AGN which are 
    thought to have relativistic jets oriented at a small angle to the line 
    of sight, so that the observed flux is strongly enhanced by relativistic 
    beaming effects.
    Detailed observations since 1997 have confirmed the picture of powerful
    activity in Mkn~501 and revealed a wide range of variability patterns 
    during outbursts (e.g. Protheroe et al.~\cite{protheroeetal1998}) 
    including flaring episodes of several days and rapid variability on 
    subhour timescale (e.g. Sambruna et al~\cite{sambrunaetal2000}). 
    The evidences for a $23$-day periodicity in both the TeV and the X-ray 
    lightcurves during the 1997 high state (cf. Hayashida et 
    al.~\cite{hayashidaetal1998}; Kranich et al.~\cite{kranichetal1999}; 
    Nishikawa et al.~\cite{nishikawaetal1999}) may be counted among the 
    most fascinating features (see Kranich~\cite{kranichdiss2001} and
    Kranich et al.~\cite{kranichetal2001} for a recent assessment of 
    significance) and possibly indicates the presence of a BBHS in Mkn~501 
    (cf. Rieger \& Mannheim~\cite{riegermannheim2000}).
    In this paper we analyse the implications of the apparent disagreement 
    of independent central mass estimates for Mkn~501. Starting with an 
    investigation of the constraints given by host galaxy observations 
    (Sect.~2), we proceed with an analysis of the robustness of the HE 
    emission estimates (Sect.~3) and finally consider the consequences 
    expected from a binary BH modelling (Sect.~4).

\section{Central mass estimates derived from host galaxy observations} 
   Recent mass estimates using host galaxy observations mainly rely on 
   the following two statistical correlations:\\
   (1) Dynamical studies of nearby elliptical galaxies have revealed an 
   apparent, almost linear correlation (albeit with significant intrinsic 
   scatter) between the central BH mass and the B-band luminosity of the 
   bulge part of the host galaxy, which is given by $M_H \simeq 0.78 \times 
   10^8 M_{\odot} (L_{\rm B,bulge}/10^{10} L_{\odot})^{1.08}$ (Magorrian 
   et al.~\cite{magorrianetal1998}; Kormendy \& Gebhardt~\cite{KG2001}, 
   hereafter KG01).\\
   (2) A much tighter correlation $M_H \propto \sigma^{\alpha}$ seems to 
   exist between the BH mass $M_H$ in nearby inactive galaxies and the 
   stellar velocity dispersion $\sigma $ of their host bulge (Gebhardt et 
   al.~\cite{gebhardtetal2000}; Ferrarese \& Merritt~\cite{FM00}). However,
   up to now there is considerable debate over the true slope $\alpha$. 
   Using different samples, Gebhardt et al.~(\cite{gebhardtetal2000})
   found  $\alpha=3.75 \pm 0.3$, while Ferrarese \& Merritt~(\cite{FM00}) 
   obtained $\alpha=4.8 \pm 0.54$ and $\alpha=4.72 \pm 0.36$ (Merritt \& 
   Ferrarese~\cite{MF01b}).
   Currently, further research is required to settle the question 
   whether this difference is mostly caused by lower quality data and a 
   less precise regression algorithm (cf. Merritt \& Ferrarese~\cite{MF01b}) 
   or by systematic differences in the velocity dispersions used by the 
   different groups for the same galaxies (cf. Tremaine et 
   al.~\cite{tremaineetal2002}).\\
   The results using reverberation mapping (RM) (e.g. Kaspi et 
   al.~\cite{kaspietal2000}; Nelson~\cite{nelson2000}; 
   Wandel~\cite{wandel2002}) indicate that the $M_H -\sigma$ correlation 
   may also hold for nearby AGN. However, a critical test of this conclusion 
   depends on both a secure measure of the BH mass and an accurate 
   determination of the stellar velocity dispersion. So far, the quality of 
   BH mass estimates from stellar or gas kinematical data (whether ground- or
   HST-based), which require the BH sphere of influence $r_H=G M_H/\sigma^2$ 
   to be well-resolved, seems to increase only modestly (Merritt \& 
   Ferrarese~\cite{MF01-ASP}), so that over-estimation may be quite possible.
   Further progress has been expected using RM methods (Ferrarese et 
   al.~\cite{ferrareseetal2001}). Yet, the accuracy of RM may be strongly 
   affected by systematic errors, e.g. due to uncertainties in the geometry 
   and kinematics of the BLR or due to an unknown angular radiation pattern 
   of the line emission, which may result in a systematic error up to at 
   least a factor of $3$ (cf. Krolik~\cite{krolik2001}). Moreover, only few 
   accurate measurements of $\sigma$ seem to exist for AGN. Ferrarese et 
   al.~(\cite{ferrareseetal2001}) have recently analysed six AGN with 
   well-determined RM BH masses by a careful measurement of their velocity 
   dispersions and found a general consistency with the $M_H - \sigma$ 
   relation for quiescent galaxies. However, only BH masses below $\simeq 
   10^8\,M_{\odot}$ have been included so far, thus leaving out the high 
   mass end of the correlation, and in addition, a large scatter is indicated. 
   Besides providing a promising tool for the determination of BH masses in 
   AGN, the current uncertainties in the correlations should be considered, 
   if one tries to assess its implication for individual sources such as
   Mkn~501.

   In the case of Mkn~501, Barth et al.~(\cite{barthetal2002a}) have 
   recently determined the stellar velocity dispersion from the calcium 
   triplet lines to be $\sigma =(372\pm18)$ km/s (cf. also Barth et 
   al.~\cite{barthetal2002b}). Applying the $M_H -\sigma$ relations of 
   KG01 and Merritt \& Ferrarese~(\cite{MF01b}), they derived a BH mass 
   for Mkn~501 of $M_H \simeq (0.9-3.4) \times 10^9\,M_{\odot}$. 
   This mass estimate was supported by the study of Wu et 
   al.~(\cite{wuetal2002}), who estimated the velocity dispersions and BH 
   masses from the fundamental plane for ellipticals for a large AGN sample 
   including $63$ BL Lac objects (but not Mkn~501). They derived BH masses 
   up to $10^9\, M_{\odot}$, but with a potential error up to a factor of two. 
   In particular, inspection of the fit in their Fig.~1 indicates a possible
   BH mass for Mkn~501 of $\sim (4-7) \times 10^8\,M_{\odot}$ for 
   $M_R({\rm host})=-24.2$ mag (Pursimo et al.~\cite{pursimoetal2002}). 
   The general challenge of determining $\sigma$ accurately may be 
   illustrated in more detail with reference to the recent work by Falomo 
   et al.~(\cite{falomoetal2002}), who provided a systematical 
   study of the stellar velocity dispersion in seven BL Lacs. Using 
   measurements in two spectral ranges, they found a velocity dispersion 
   of $\sigma=(291\pm 13)$ km/s for Mkn~501, which is significantly lower 
   than the one derived by Barth et al.~(\cite{barthetal2002a}). Hence, if 
   this value is used instead, the BH mass estimated by Barth et 
   al.~(\cite{barthetal2002a}) is reduced by up to a factor of three, i.e. 
   one obtains $M_H \simeq (3.6 - 10.7) \times 10^8\, M_{\odot}$. Additional 
   support for such a low $\sigma$-value in Mkn~501 seems to be indicated by 
   the original Faber \& Jackson relation, which yields $\sigma \sim 270$ km/s
   (see Fig.~2 in Falomo et al.~\cite{falomoetal2002}) for $M_R({\rm host})=
   -24.2$ mag. Future research is needed to test whether the discrepancy
   in $\sigma$ is mainly induced by the difference in the method deriving 
   $\sigma$ (direct fitting versus Fourier quotient routine).
   
   As noted above, an additional mass estimate for Mkn~501 can also be 
   derived from the $M_H-L_{\rm bulge}$ correlation. The reported large 
   uncertainties in this relation have recently been examined by McLure \& 
   Dunlop~(\cite{mclure2002}) using R-band luminosities, which are less 
   sensitive to extinction. By analysing the virial BH masses for a sample 
   of $72$ AGN, they found the scatter to be quite smaller than previously 
   estimated and stressed its usefulness. For application to Mkn~501,
   we may exploit the absolute R-band luminosity of its host galaxy 
   recently derived by Pursimo et al.~(\cite{pursimoetal2002}) 
   (see also Nilsson et al.~\cite{nilssonetal1999}). Assuming $H_0=50$ 
   km s$^{-1}$ Mpc$^{-1}$, they obtained $M_R({\rm host}) = -24.2$ mag. 
   If we convert R- to B-band luminosity assuming $B-R=1.56$ (e.g. Goudfrooij 
   et al.~\cite{goudfrooijetal1994}; Fukugita et al.~\cite{fukugitaetal1995}; 
   Urry et al.~\cite{urryetal2000}), we have $M_B({\rm host})=-22.46$ mag, 
   which results in $L(M_B({\rm host}))=1.406 \times 10^{11} L_{\odot}$. 
   Using the KG01-relation for the B-band luminosity, the expected BH mass 
   in Mkn~501 is $M_H \simeq 1.3 \times 10^9 \,M_{\odot}$, but with a 
   potential error of up to at least a factor of three. 
   Using the more recent McLure \& Dunlop~(\cite{mclure2002})-relation 
   $\log(M_H/M_{\odot})=-0.5 M_R - 2.96 (\pm 0.48)$, one finds $M_H \simeq 
   (0.46-4.2) \times 10^9\,M_{\odot}$. Uncertainties in the determination of 
   $M_R$ may further reduce the expected BH mass. For example, values from the
   literature presented in Table~4 of Nilsson et al.~(\cite{nilssonetal1999}) 
   indicate that $M_R$ might be up to $0.4$ mag higher and therefore $M_H$ 
   correspondingly smaller. More importantly, if a Hubble constant $H_0=70$ 
   km s$^{-1}$Mpc$^{-1}$ is assumed, one finds $M_R=-23.47$, which results 
   in $M_H \simeq 7.8 \times 10^8 \,M_{\odot}$ (KG01), again with substantial 
   scatter of up to at least a factor of three, or $M_H \simeq (2-18) \times 
   10^8 \,M_{\odot}$  (McLure \& Dunlop~\cite{mclure2002}), thus allowing for 
   a central mass as low as $2 \times 10^8 \,M_{\odot}$.

\section{Mass estimates derived in the context of high energy emission models}
   Following a quite different approach, the BH mass of Mkn~501 could also 
   be estimated in the framework of high energy emission models:\\ 
   (1) With respect to the high energy emission, Fan et al.~(\cite{fan99}) 
   have recently determined the central black hole masses for several 
   $\gamma$-ray loud blazars (including Mkn~501) by assuming that the 
   observed $\gamma$-rays are produced at $\simeq 200\,R_g$.
   Accordingly, the central BH mass obeys the relation $M_H/M_{\odot}\simeq 
   500\,\delta\, \Delta t$, where $\Delta t$ denotes the doubling timescale 
   in seconds and $\delta$ the Doppler factor. For Mkn~501 they obtained a 
   central BH mass of $M_H \simeq 0.9 \times 10^7 M_{\odot}$, using $\Delta t 
   =6$ hrs and $\delta \simeq 0.9$. 
   However, current evidence indicates that the relevant timescale $\Delta t$ 
   might be substantially smaller. Recent observations reveal considerable 
   sub-hour variability (e.g. Ghosh et al.~\cite{ghoshetal2000}; Catanese \& 
   Sambruna~\cite{catanesesambruna}) on a timescale as low as $\Delta t = 
   1200$ s (Sambruna et al.~\cite{sambrunaetal2000}). Such a low value 
   for the observed timescale could possibly be accommodated by assuming a 
   high doppler boosting factor $\delta \simeq 18$. Hence, it appears that 
   the more crucial point in this derivation is the assumption that the 
   $\gamma$-rays dominating the emission are produced at $\sim 200$ 
   Schwarzschild radii. Indeed, at least in the case of the TeV-blazars, 
   the variable, high energy emission is usually regarded as produced by 
   moving knots or shocks in the jet far from the accretion disk (for a 
   review, cf. Mannheim~\cite{mannheim97}; Aharonian \& 
   V\"olk~\cite{ahavoelk2001} and references therein). 
   While instructive, the derived estimate should thus not be 
   considered as a robust, general limit.\\
   (2) A further mass estimate for Mkn~501 has been derived by DeJager et 
   al.~(\cite{dejager1999}) following an approach developed by Hayashida \& 
   Miyamoto et al.~(\cite{hayashidaetal1998k}). Assuming the variation in the
   accretion process to drive the X-ray and TeV variation in the jet via the 
   dynamo effect, their result yields a central BH mass of $M_H=(1-6) \times 
   10^7\,M_{\odot}$ for $\delta=(10-15)$. However, due to the absence of a 
   physical basis for the required scaling of the Fourier spectrum and due
   to the assumption of a linear proportionality between variability timescale 
   and BH mass (cf. Hayashida \& Miyamoto et al.~\cite{hayashidaetal1998k}), 
   which is probably not valid for the blazar class (Kataoka et 
   al.~\cite{kataokaetal2001}), this estimate again does not appear to be 
   robust.\\
   (3) The estimates (1) and (2) which suggest a BH mass less than $\sim 6
   \times 10^7\,M_{\odot}$, are strongly model-dependent as shown above.
   We may illustrate this in more detail by comparing them with results derived
   in the context of another, high energy emission model. Bednarek, Kirk \& 
   Mastichiadis~(\cite{bednareketal1996}) for example, have developed a 
   special model for the origin of the high energy particles in TeV blazars 
   like Mkn~421 and Mkn~501, assuming the electrons responsible for the high 
   energy emission to be accelerated rectilinearly in an electric field. 
   In this model, the mass of the central BH is expected to be limited by $M_H 
   \gppr 10^8 M_{\odot}\,(E_{\rm max}/34\, {\rm TeV})^{2.5}/l_{\rm Edd}^{1/2}$,
   where $E_{\rm max}$ denotes the maximum photon energy and $l_{\rm Edd}$ the
   disk luminosity in units of the Eddington luminosity. There is strong 
   evidence for a sub-Eddington accretion mode in BL Lacs in general (e.g. 
   Cavaliere \& D'Elia~\cite{cavalieredelia2002}) and particularly
   for the TeV emitting blazars (e.g. Celotti et al.~\cite{celottietal1998}).
   Thus, using characteristic values, i.e. $l_{\rm Edd}=(0.01-0.001)$ and
   $E_{\rm max}=20$ TeV (cf. Samuelson et al.~\cite{samuelsonetal98}; 
   Konopelko~\cite{konopelko99}), we arrive at a mass $M_H \gppr (2.46-14.66)
   \times 10^8\, M_{\odot}$, which is up to ten times larger than the
   estimates (1) and (2).

\section{Mass estimates derived in the context of a binary black hole scenario}
\begin{figure}
   \centering
   \includegraphics[width=7cm]{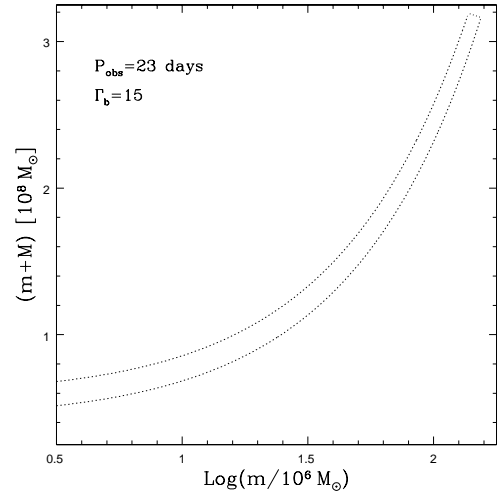}
      \caption{Allowed central mass dependence as a function of the secondary
       mass assuming a BBHS, where the observed periodicity of $23$ days is 
       related to the orbital motion of a jet, which emerges from the less 
       massive BH. The allowed mass range lies inside the curves and is 
       calculated using $\Gamma_b=15$.} 
      \label{kepler}
   \end{figure}

    In a recent contribution (Rieger \& Mannheim~\cite{riegermannheim2000},
    hereafter RMI; \cite{riegermannheim2001}), we have shown that the 
    periodicity of $P_{\rm obs}=23$ days, observed during the 1997 high state 
    of Mkn~501, could be plausibly related to the orbital motion in a BBHS, 
    provided the jet, which dominates the observed emission, emerges from the 
    less massive (secondary) BH. 
    If such an interpretation (henceforth called the standard scenario) is 
    appropriate, we may derive a third estimate for the central mass in 
    Mkn~501.     
    We have demonstrated in RMI that, due to relativistic effects, the 
    observed period appears drastically shortened, so that for the intrinsic 
    keplerian orbital period one finds $P_k = (6-14)\,\,{\rm yrs}\,$.
    Taking into account that the observed emission is periodically modulated 
    by differential doppler boosting due to the orbital motion, one may derive 
    a simple equation for the required mass dependence in the standard scenario
    (cf. RMI, Eq.~(8)):
    \begin{eqnarray}\label{ratio1}
    \frac{M}{(m+M)^{2/3}} & = & \frac{P_{\rm obs}^{1/3}}{(2\,\pi\,[1+z]\,
                    G)^{1/3}}\,\frac{c}{\sin i}\,
                    \nonumber \\            
                    & &\times \frac{f^{1/(3+\alpha)}-1}{f^{1/(3+\alpha)}+1}\,
                        (1-\frac{v_z}{c}\,\cos i)^{2/3}\,.
    \end{eqnarray} Here $\alpha$ denotes the spectral flux index, $z$ the 
    redshift, $v_z$ the outflow velocity, $i\simeq 1/\Gamma_b $ the angle of 
    the jet axis to the line of sight, $\Gamma_b \simeq 10-15$ (e.g. 
    Mannheim et al.~\cite{mannheimetal1996}; Spada et al.~\cite{spadaetal1999})
    the bulk Lorentz factor and $f$ the observed flux ratio between maximum 
    and minimum (for the TeV range $f \sim 8$), while $m$ and $M$ denote the 
    masses of the smaller and larger BH, respectively.\\
    In order to break the degeneracy in this mass ratio, we may utilize an 
    additional constraint by assuming that the current binary separation $d$ 
    corresponds to the separation at which gravitational radiation becomes 
    dominant (cf. RMI). Such a constraint yields an upper limit for the 
    allowed binary masses and might be associated with the key aspect that 
    BL Lac objects are old, more evolved and underluminous sources, i.e. 
    they might be close binaries, probably settled above or near the critical 
    gravitational separation, because the possibility of removing further 
    angular momentum has been almost terminated as a result of declining
    gas accretion rates.
    We can specify the corresponding gravitational separation $d_g$ by 
    equating  the timescale $\tau_{\rm grav}=|d/\dot{d}| = (5\, c^5/64\, G^3)
    \,d^4/(M\,m\,[m+M])$ on which gravitational radiation shrinks the binary 
    orbit, with the dynamical timescale $\tau_{\rm gas}$ for gas accretion 
    (cf. Begelman et al.~\cite{begeletal80}; note that compared with RMI,
    this estimate for $\tau_{grav}$ is a factor $2.5$ more precise, 
    cf. Rieger \& Mannheim~\cite{riegermannheim2001}).
    A characteristic measure for  $\tau_{\rm gas}$ is given by the Eddington 
    limit $\tau_{\rm gas}\simeq 3.77 \times 10^7\,(\eta/0.1)$ yrs, assuming a 
    canonical $10\%$ conversion efficiency (cf. Krolik~\cite{krolik99}).
    Using $\tau_{\rm grav}$ and $\tau_{\rm gas}$, one finally arrives 
    at 
    \beq\label{grav-distance}
       d_g \simeq 3.50 \times 10^{16}\,M_8^{1/4}\,m_8^{1/4}\, (m_8+M_8)^{1/4}
              \,\,\,{\rm cm}\,.
    \eeq
    If we combine this expression with the relevant expression for the binary 
    separation (i.e. Eq.~(7) of RMI), we obtain a second constraint 
    on the allowed mass ratio given by
    \beqn\label{ratio2}
      \frac{(m_8 + M_8)^{3/4}}{M_8^{5/4}\,m_8^{1/4}}
                   = 7.3 \times 10^{6}\,
                     \frac{f^{1/(3+\alpha)}+1}{f^{1/(3+\alpha)}-1}\,
                     \frac{(1+z)}{P_{\rm obs}}\,\sin i\,,
    \eeqn assuming the jet to arise from the less massive BH (for the
    reverse case the BH masses should be interchanged in Eq.~(\ref{ratio1}) 
    and Eq.~(\ref{ratio2})). By using Eq.~(\ref{ratio1}) and 
    Eq.~(\ref{ratio2}), we may determine the binary masses permitted for 
    a BBHS with separation near that for which gravitational radiation 
    becomes dominant. 
    In Fig.~\ref{kepler} we have plotted the allowed range for the case of 
    $\Gamma_b=15$ and $\alpha \simeq (1.2-1.7)$ (e.g. Aharonian et 
    al.~\cite{aharonianetal1999}). The run of the curves is determined by 
    Eq.~(\ref{ratio1}) with the upper bounds given by Eq.~(\ref{ratio2}). 
   We note that the upper bounds depend on the presumed timescale for the 
   binary evolution, i.e. on the estimate of the gravitational separation. 
   In particular, the derived upper bounds may be larger, if, for example, 
   disk-driven migration (e.g. Armitage \& Natarajan~\cite{armitage2002}) 
   might be relevant. Fig.~\ref{kepler} reveals that (a) the typical combined 
   mass for a BBHS is expected to be less than $\sim 3 \times 10^8 M_{\odot}$,
   and that (b) a BBHS with, e.g. $m \sim 6 \times 10^7 M_{\odot}$ and 
   combined central mass $(M+m) \simeq (1.5-2) \times 10^8 M_{\odot}$ appears 
   well-conceivable given the current limits.\\
   On the other hand, if a high central mass of $\sim 10^9\,M_{\odot}$ will
   be established by further research, the proposed binary scenario appears
   to be ruled out. We note, however, that even in this case a binary scenario 
   may be still possible provided that the jet, which dominates the emission, 
   is produced by the primary BH. To illustrate the implications in this case,
   let us consider a (combined) central mass of $\simeq 0.9 \times \times 
   10^9\,M_{\odot}$ by demanding the primary to be in the range $(5-6) \times 
   10^8\,M_{\odot}$ (see Table~\ref{table2}). 
   The mass of the secondary BH then is determined by Eq.~(\ref{ratio1}) with 
   the masses interchanged, the current separation $d$ by Eq.~(3) of RMI 
   (again with the masses interchanged) and the gravitational separation $d_g$
   by Eq.~(\ref{grav-distance}).  
   The results as shown in Table~\ref{table2} imply a close binary system with 
   $d \lppr d_g$. If the optically bright QSO stage thus occurs during the 
   binary evolution and the applied doppler factors are considered as 
   typical, phases of super-Eddington accretion and/or with decreased 
   conversion efficiency seem to be necessary for the binary to be above 
   its gravitational separation.  
  \begin{table}[hbt]
  \vspace{-0.2cm}
  \begin{center}
  \begin{tabular}{|l@{\hspace{0.5cm}}|l@{\hspace{0.5cm}}
                |l@{\hspace{0.5cm}}|}\hline
  $i=1/\Gamma_b$  &  $1/10$       &  $1/15$  \\ \hline \hline
  $m\,[10^8\,M_{\sun}]$   & 4.22 \quad (3.62) & 3.85 \quad (3.34) \\ \hline
  $M\,[10^8\,M_{\sun}]$   & 5.00 \quad (5.00)   & 6.00 \quad (6.00) \\ \hline
  $d\,[10^{16}\,\rm{cm}]$ & 4.87 \quad (4.76)   & 8.54 \quad (8.39) \\ \hline
  $d_g\,[10^{16}\,\rm{cm}]$ &13.1 \quad (12.4)   & 13.6 \quad (12.9) \\ \hline
  \end{tabular}
  \end{center}
  \vspace{0cm}
  \caption{Binary masses for a central core mass of $\simeq 0.9 \times 10^9 \,
           M_{\odot}$, currently expected separation $d$ and corresponding 
           gravitational separation $d_g$. The calculations have been done 
           using $\alpha=1.2$ (in brackets: $\alpha=1.7$). The jet is assumed 
           to be produced by the more massive BH.}\label{table2}
  \vspace{-0.2cm}
  \end{table}
\vspace{-0.3cm}

\section{Discussion and conclusions}
   In this paper, we have considered the apparent discrepancies between 
   independent central mass estimates for the paradigmatic object Mkn~501.
   Our main conclusions can be summarized as follows:\\
   (1) Estimates inferred from host galaxy observations using the reported, 
   statistical correlations for the central velocity dispersion and the bulge 
   luminosity reveal a tendency for the central BH mass in Mkn~501 to be
   larger than $\sim 5 \times 10^8\,M_{\odot}$. If verified by further 
   investigations, this would exclude both the standard binary scenario, where 
   the jet is emitted from the less massive BH, and several high energy 
   emission models proposed so far. However, as the correlations still show 
   substantial intrinsic scatter and large uncertainties due to the use of 
   limited samples -- especially with respect to the high mass end -- the 
   accuracy of this estimate should be treated with caution. In particular, 
   we have shown that the central mass limit could be easily up to a factor of 
   $(2-3)$ smaller, thus allowing for the standard binary scenario, in 
   which case a simple explanation for the divergent mass estimates from 
   emission models and host observations appears possible.\\
   (2) Independent mass estimates from high energy emission models suggest
   a mass for the jet emitting BH in Mkn~501 which is smaller than $\simeq 6 
   \times 10^7\, M_{\odot}$. However, this estimate is quite model-dependent
   and so cannot be used as a universally valid constraint. In particular, 
   emission models have been developed for Mkn~501 where a central BH mass 
   up to ten times larger appears quite possible.\\ 
   (3) If a binary BH interpretation for the periodical variability (cf. RMI)
   is appropriate in the case of Mkn~501, the jet dominating the emission
   has to be produced by the less massive (secondary) BH. 
   Using characteristic jet parameters and assuming both Eddington-limited 
   accretion and a binary separation comparable to the gravitational one, 
   we have shown that the combined central (primary $+$ secondary BH) 
   mass should be smaller than $\simeq 3 \times 10^8\,M_{\odot}$.
   The binary model particularly permits a system with $m=6 \times 10^7
   M_{\odot}$ and $(m+M)\simeq (1.5-2)\times 10^8\,M_{\odot}$, possibly
   allowing for a convergence of constraints from emission models and host 
   observations.\\
   (4) If further research clearly establishes a high central BH mass $\sim 
   10^9 \,M_{\odot}$ in Mkn~501, this would call for a change of fundamental 
   assumptions and/or a modification of the parameter space regarded 
   to be typical in the context of several high energy emission models. 
   The only way to avoid this conclusion seems to be the assumption of a 
   highly unequal BBHS with the jet dominating the emission produced by 
   the less massive and the central mass dominated by the primary BH. 
   However, it then appears no longer be possible to explain the observed 
   periodicity via the orbital motion. On the other hand, even in the case 
   of a high central mass, a binary model for the observed periodicity may 
   be still possible provided the jet is produced by the primary BH. 
   Yet, for the binary to be near or above the separation at 
   which gravitational radiation becomes dominant, one then requires the 
   earlier binary evolution to pass through phases of super-Eddington 
   accretion and/or decreased conversion efficiency $\eta < 0.1$.
   Such conditions need not necessarily be ad hoc but have already been 
   considered in the context of galaxy evolution and the growth of massive 
   BH (e.g. Collin et al.~\cite{collinetal2002}; Yu \& 
   Tremaine~\cite{yutremaine2002}).\\
   In view of Mkn~501 as a paradigmatic object, bringing its mass 
   estimates to convergence continues to be an important task. Further 
   research on the parameter space of emission models and the careful 
   observations of nearby galaxies will be particularly valuable in order 
   to minimise the uncertainties in the statistical correlations, to evaluate 
   the impact of host observations on high energy emission models, and to 
   assess the plausibility of a BBH model.
\begin{acknowledgements}
      We are grateful to J. Heidt for fruitful discussions about 
      the host of Mkn~501, and C. Hettlage and R. Hessman for a 
      helpful reading of the manuscript. F.M.R. gratefully acknowledges 
      support under DFG MA 1545/8-1.
\end{acknowledgements}

\end{document}